\documentclass[12pt]{article}
\usepackage{tabularx}
\usepackage{array}
\usepackage{graphics}
\usepackage{epsfig}
\usepackage{amssymb}
\usepackage{citesort}
\usepackage{colortbl}
\usepackage{pstricks}
\usepackage{hhline}
\usepackage{multirow}

\makeatletter

\usepackage{verbatim}

\newcommand{\msbar}{\overline{\rm MS}}
\newcommand{\bea}{\begin{eqnarray}}
\newcommand{\eea}{\end{eqnarray}}
\newcommand{\beq}{\begin{equation}}
\newcommand{\eeq}{\end{equation}}
\newcommand{\gev}{{\rm GeV}}

\newcommand{\pdir}{p\kern -5.2pt\raise 0.2ex\hbox {/}}
\newcommand{\vdir}{v\kern -5.75pt\raise 0.15ex\hbox {/}}
\newcommand{\kdir}{k\kern -5.75pt\raise 0.15ex\hbox {/}}
\newcommand{\epsdir}{\epsilon\kern -5.0pt\raise 0.15ex\hbox {/}}
\newcommand{\bvdir}{\bar{v}\kern -5.75pt\raise 0.15ex\hbox {/}}
\newcommand{\Ddir}{D\kern -7.75pt\raise 0.20ex\hbox {/}}
\newcommand{\ldir}{l\kern -5.0pt\raise 0.2ex\hbox{/}}
\newcommand{\varepsdir}{\varepsilon\kern -5.5pt\raise 0.15ex\hbox{/}}

\newcommand{\bbbard}{\overline B_d^0-B_d^0}
\newcommand{\bbbars}{\overline B_s^0-B_s^0}

\def\ndot{\!\cdot\! }
\def\bkg{$B \rightarrow K^\ast \gamma\ $}

\def\ds{\displaystyle}
\definecolor{grisfonce}{rgb}{0.8,0.8,0.8}
\definecolor{grisclaire}{rgb}{0.95,0.95,0.95}

\makeatother

\begin{document}
\thispagestyle{empty} 
\begin{flushright}
\begin{tabular}{l}
LPT - Orsay 06/73\\
RM3-TH/06-22
\end{tabular}
\end{flushright}
\begin{center}
\vskip 3.0cm\par
{\par\centering \textbf{\LARGE  
An estimate of the \bkg}}\\
\vskip 0.3cm\par
{\par\centering \textbf{\LARGE   
decay form factor}}\\
\vskip 1.25cm\par
{\scalebox{.9}{\par\centering \large  
\sc Damir Be\'cirevi\'c$^a$, Vittorio Lubicz$^b$ and Federico Mescia$^{c}$}}
{\par\centering \vskip 0.75 cm\par}
{\sl 
$^a$ Laboratoire de Physique Th\'eorique (B\^at.~210),\\
Universit\'e Paris Sud, Centre d'Orsay,\\ 
F-91405 Orsay-Cedex, France. \\
\vspace{.25cm}
$^b$ Dip. di Fisica, Univ. di Roma Tre and INFN,
Sezione di Roma III, \\
Via della Vasca Navale 84, I-00146 Rome, Italy.\\
\vspace{.25cm}
$^c$ INFN, Laboratori Nazionali di Frascati,\\
Via E. Fermi 40, I-00044 Frascati, Italy.}\\
\vskip1.cm
 
{\vskip 0.35cm\par}
\end{center}

\vskip 0.55cm
\begin{abstract}
We present the results of a lattice QCD calculation 
of the form factor relevant to \bkg decay. 
Our final value, $T(0)=0.24\pm 0.03^{+0.04}_{-0.01}$, is obtained in 
the quenched approximation, and by extrapolating  
$M_H^{3/2}\times T^{H\to K^\ast}(q^2=0)$ from the directly accessed 
$H$-heavy mesons to the meson $B$. We also show that the extrapolation from 
$B \rightarrow K^\ast \gamma^\ast$ ($q^2\neq 0$) 
to $B \rightarrow K^\ast \gamma$ ($q^2= 0$), leads to a result 
con\-sis\-tent with the one quoted above. 
On the other hand, our results are 
not accurate enough to solve the $SU(3)$ flavor breaking effects in the 
form factor and we quote 
$T^{B\to K^\ast}(0)/T^{B\to \rho}(0)=1.2(1)$, as our best estimate.
\end{abstract}
\vskip 1.6cm
{\small PACS: 12.38.Gc,\ 13.25.Hw,\ 13.25.Jx,\ 13.30.Ce,\ 13.75Lb} 
\vskip 2.2 cm 
\setcounter{page}{1}
\setcounter{footnote}{0}
\setcounter{equation}{0}
%%%%%%%%%%%%%%%%%%%%%%%%%%%%%%%%%%%%%%%%
%%%%%%%%%%%%%%%%%%%%%%%%%%%%%%%%%%%%%%%%
%%%%%%%%%%%%%%%%%%%%%%%%%%%%%%%%%%%%%%%%
\noindent

\renewcommand{\thefootnote}{\arabic{footnote}}

\newpage
\setcounter{footnote}{0}
%%%%%%%%%%%  Section 1

\section{Introduction}
\setcounter{equation}{0}

The flavour changing neutral decays,  
$B\to V\gamma$ ($V=K^\ast, \rho, \omega$), are induced by  penguin diagrams.  
Their accurate experimental measurement gives us information about the  
heavy particle content in the loops, and thus might be a window
to the physics beyond the Standard Model (SM). This is why a huge 
amount of both experimental and theoretical research has been invested 
in studying these modes over the past decade.

The experimenters at CLEO were first to observe and measure the \bkg decay 
rate~\cite{Ammar}. Averaging over the neutral and charged $B$-mesons they reported
\bea
B\left(B \to K^{\ast}\gamma\right) = (4.5\pm 1.5\pm 0.9)\times 10^{-5}.
\eea
Today, the unprecedented statistical quality of the data collected at 
the $B$-factories made it possible to do precision measurements separately for $B^0$ and $B^\pm$ decays, namely
\bea\label{eq1}
B\left(B^0 \to K^{\ast 0}\gamma\right)&=&
\left\{
\begin{array}{cc}
(3.92\pm 0.20\pm 0.24) \times 10^{-5}&{\rm BaBar}~\cite{babar}  \\
\hfill \\
(4.01\pm 0.21\pm 0.17) \times 10^{-5}&{\rm Belle}~\cite{belle} \\
\end{array}
\right. \;,\nonumber \\
\hfill  \\
B\left(B^+ \to K^{\ast +}\gamma\right)&=&
\left\{
\begin{array}{cc}
(3.87\pm 0.28\pm 0.26) \times 10^{-5}&{\rm BaBar}~\cite{babar}  \\
\hfill\nonumber \\
(4.25\pm 0.31\pm 0.24) \times 10^{-5}&{\rm Belle}~\cite{belle} \\
\end{array}
\right. \nonumber\;.
\eea
Besides, the first significant measurements of 
$B\to \rho(\omega)\gamma$~\cite{brhoEXP}, opened a discussion 
on the possibility of constraining $\vert V_{td}/V_{ts}\vert$, thus providing 
an alternative to the constraint arising from the ratio of the oscillation
frequencies in the $\bbbars$ and  $\bbbard$ systems, $\Delta m_{B_s}/\Delta m_{B_d}$~\cite{gerhard}.

However, when looking for non-SM effects in these decays,
one should be able to confront the above experimental results to the 
corresponding theoretical estimates within the SM.~\footnote{A review 
on rare $B$-decays, containing an extensive list of references can be found in~\cite{Hurth}.} 
As usual, the main obstacle is a lack of good theoretical control over 
the hadronic uncertainties. The hadronic matrix element entering the analysis 
of these, electromagnetic penguin induced, decays is
\bea
&&\hspace*{-22mm}\langle V(p^\prime;e_\lambda )\vert T^{\mu \nu}(0) \vert B(p)\rangle
= e^{\ast}_\alpha(p^\prime,\lambda) \times {\cal T}^{\alpha\mu\nu }\,,\cr 
&&\cr
&&\cr
 &&\hspace*{-7mm}{\cal T}^{\alpha\mu\nu }= \epsilon^{ \alpha\mu \nu \beta} 
\left[
\left(  p_\beta + p^{\prime}_\beta - {M_B^2 - m_V^2 \over q^2 } q_\beta \right) T_1(q^2)
+ {M_B^2 - m_V^2 \over q^2 } q_\beta 
 T_2(q^2) \right]  \cr
&&\hspace*{8mm} + {2 p^\alpha \over q^2}  \epsilon^{\mu \nu \sigma \lambda} 
p_\sigma p^{\prime}_{\lambda} \left( T_2(q^2)- T_1(q^2)
 + {q^2\over M_B^2-m_V^2}T_3(q^2) \right) \,,
\eea
where $q=p-p^\prime$, the tensor current $T^{\mu\nu}=i\overline s\sigma^{\mu\nu}b$ for $V=K^\ast$, 
and $T^{\mu\nu}=i\overline d\sigma^{\mu\nu}b$ for $V=\rho$, with $\sigma_{\mu\nu}=\frac{i}{2}[\gamma_\mu,\gamma_\nu]$. 
The above definition is suitable for the extraction of the form factors from the correlation functions computed 
on the lattice. The 
form factors $T_{1,2,3}(q^2)$ are the same as those computed 
by the QCD sum rules~\cite{LCSR,QCDSR}. 
For the physical photon ($q^2=0$), the form factors $T_1(0)=T_2(0)$, 
while the coefficient multiplying  $T_3(0)$ is zero.   

In this paper we show that the strategy which we previously employed to compute the 
$B\to \pi$ semileptonic form factors~\cite{our_BPI} can be used to compute the radiative decays 
as well. Although the attainable accuracy is quite limited, we believe the values 
we get are still phenomenologically useful. 
In what follows we will show how we obtain  
$T^{B\to K^\ast}(0)=0.24(3)\left(^{+4}_{-1}\right)$, 
from our quenched QCD calculations with ${\cal O}(a)$ improved 
Wilson quarks at two lattice spacings.  We also obtain 
$T^{B\to K^\ast}(0)/T^{B\to \rho}(0)=1.2(1)$ although that result 
is unstable when applying different strategies and using 
different lattice spacings.

\section{Methods to approaching $q^2=0$ and $M_B$}

Even though we work with ever smaller lattice spacings (``$a$"), 
we are still not able to work directly with the heavy $b$-quark. 
Instead of simulating a meson with $M_B=5.28$~GeV, 
we compute the form factors with  fictitious heavy-light 
mesons ($H$) of masses $M_B > M_H \geq M_D$, and then extrapolate them  
in $1/M_H$ to $1/M_B$, guided by the heavy quark scaling laws. 
Alternatively one can discretise the nonrelativistic QCD (NRQCD) which 
basically means the inclusion of $1/(am_b)$-corrections to the static limit.  
This, in the lattice QCD community, is  known as the {\it ``NRQCD approach"}. 
Finally, one can build an effective theory that combines the above two, which 
is known as  the {\it ``Fermilab approach"}. 
Each of the mentioned approaches has its advantages and drawbacks. 
They were all used in computing the $B\to \pi$ semileptonic 
form factors and the results show a pleasant overall agreement (see e.g. fig.3 
in the first ref.~\cite{amsterdam}).

Concerning the methodology employed while working with propagating 
heavy quarks, one should keep in mind that the form factors are accessed 
for all $q^2 \in [0, (M_H-m_V)^2]$. Only after extrapolating to $M_B$,  
at fixed values of $v\ndot p^\prime$, the $q^2$-region 
becomes large and the form factors are shifted to large $q^2$'s.~\footnote{
$v$ stands for the heavy quark (meson) four-velocity, so that $q^2 = 
M_B^2+m_V^2-2 M_B v\ndot p^\prime$. In the rest frame of the heavy meson, 
$E=v\ndot p^\prime$ is the energy of the light meson emerging from the decay.} 
The assumption underlying this extrapolation is that the HQET scaling 
laws~\cite{isgur} remain valid when $E=v\ndot p^\prime > m_V$. 
In the case of $B\to \pi\ell \nu$, it appears that this assumption is 
not particularly worrisome as the form factors, after 
extrapolating to $M_B$~\cite{our_BPI,UKQCD_BPI}, are consistent 
with those obtained by the effective heavy quark approaches~\cite{JLQCD,FNAL}, 
at large $q^2$'s. 
If one is interested in the form factor at $q^2=0$, then 
in order to extrapolate from large $q^2$'s one has to make 
some physically motivated assumption about the $q^2$-shapes 
of the form factors.

Otherwise, when working with propagating heavy quarks, one can extrapolate 
the form factors directly 
computed on the lattice at $q^2=0$ in $1/M_H$ to $1/M_B$ . 
The useful scaling law relevant 
to this situation was first noted in the framework of the light cone QCD sum rules 
(LCSR)~\cite{chernyak}, then generalised to the large energy effective theory in ref.~\cite{jerome}, and 
 finally confirmed in the soft collinear effective theory~\cite{SCET}. 
 The underlying assumption in this extrapolation is that the scaling law would 
 remain valid even when the light meson is not very energetic 
 (in the rest frame of the heavy, the $q^2=0$ point corresponds 
 to $E=(M_H^2+m_V^2)/2M_H$).

 In ref.~\cite{our_BPI} we showed that the results for the $B\to \pi \ell\nu$ form factor 
 at $q^2=0$, obtained by employing either of these two different ways of extrapolating 
 to $M_B$, are fully compatible. 
 The method of extrapolating in $1/M_H$ at $q^2=0$ fixed is  
 particularly useful for $B\to K^\ast(\rho)\gamma$, where the main goal is 
 to compute the form factor when the photon is on-shell, $T(q^2=0)$. 
 This is what we do in this paper. As a cross-check of our result 
 we also use the standard method (extrapolating in $1/M_H$ prior to the extrapolation  
 in $q^2$, down to $q^2=0$).

\section{Raw Lattice Results}

The form factors are extracted from the study of suitable ratios
of three- and two-point correlation functions, namely
\bea
\label{ratio}
R_{\zeta \mu\nu}(t_y) = 
{ {\cal C}^{(3)}_{\zeta \mu\nu } (t_x,t_y;\vec q, \vec p_H)\over \quad \frac{1}{3}\ds{
\sum_{\alpha=0}^3} {\cal C}^{(2)}_{\alpha\alpha}(t_y,\vec p^{\ \prime}) 
\, \ {\cal C}^{(2)}_{HH}(t_x - t_y,\vec p_H)  \quad } \times 
{ \sqrt{{\cal Z}_{H}} \sqrt{{\cal Z}_{V}} }\;,
\eea
where $\vec p^{\ \prime}=\vec p_H - \vec q$. The correlation functions and their asymptotic 
behavior are given by
\bea \label{asym3}
&&{\cal C}_{HH}^{(2)}(t,\vec p) = \sum_{\vec x} \ e^{i\vec p\cdot \vec x } 
 \langle   P_5({\vec x}, t) 
P_5^{\dagger}(0) \rangle \stackrel{t\gg 0}{\longrightarrow}  {{\cal{Z}}_H
 \over 2 E_H}e^{- E_H t} \,,\cr
 &&\cr
 &&\cr
&&{\cal C}_{\alpha\beta}^{(2)}(t,\vec p^{\ \prime}) = \sum_{\vec x} \ e^{i\vec p^{\ \prime}\cdot \vec x } 
 \langle   V_\alpha({\vec x}, t) 
V_\beta^{\dagger}(0) \rangle \stackrel{t\gg 0}{\longrightarrow} \sum_\lambda {{\cal{Z}}_V 
 \over 2 E_V} e_\alpha^\ast(p^\prime,\lambda) e_\beta(p^\prime,\lambda) e^{- E_V t} \,,\cr
&&\cr
 &&\cr
&&{\cal C}^{(3)}_{\zeta\mu\nu} (t_x, t_y;\vec q,\vec p_H)= 
\sum_{\vec x, \vec y} e^{i (\vec q\cdot \vec y - \vec p_H\cdot \vec x )} 
\langle   V_\zeta(0)\hat T_{\mu \nu}(y) P_5^\dagger(x)  \rangle  \cr 
&&\stackrel{t_x\gg t_y \gg 0}{\longrightarrow} 
\sum_\lambda {\sqrt{{\cal Z}_{V}} \over 2 E_{V}} e_\zeta^\ast(p^\prime,\lambda) \ e^{- E_{V} t_y} 
 \times 
\langle V (\vec p^{\ \prime},\lambda)\vert \hat T_{\mu,\nu}(0)
\vert H(\vec p_H)\rangle  \times {\sqrt{{\cal Z}_{H}} \over 2 E_{H}} \ e^{-
E_{H}(t_x - t_y)}\cr
&&\quad\quad=  {\sqrt{{\cal Z}_{V} {\cal Z}_{H}} \over 4 E_{V} E_H} \sum_\lambda
e_\zeta^\ast(p^\prime,\lambda)e^\alpha(p^\prime,\lambda)\, {\cal T}_{\alpha\mu\nu}\, e^{-E_{H} t_x + (E_H - E_V) t_y}.
\eea
For the interpolating field we choose $P_5=\overline q i\gamma_5 Q$ and 
$V_\mu=\overline q \gamma_\mu q$, with $q$ and $Q$ being 
light and heavy quark respectively. We also defined  
$\sqrt{{\cal Z}_{H}} = \vert\langle 0\vert  P_5\vert H \rangle\vert$, and  
$\sqrt{{\cal Z}_{V}}e_\mu^\ast(p^\prime,\lambda) = \vert\langle 0\vert  V_\mu\vert V(p^\prime,\lambda) \rangle\vert$. The hat over the tensor current indicates that it is 
${\cal O}(a)$ improved and renormalised at some scale $\mu$, i.e.,
\bea
\label{def:2}
\hat T_{\mu\nu}(\mu)= Z_T^{(0)}(g_0^2,\mu)  \, \Bigl[ 1+ b_T(g_0^2)\ \overline{a m}) \Bigr] \, 
\Bigl[ i\overline Q \sigma_{\mu\nu} q\ +\ a c_T(g_0^2) \left( \partial_\mu \overline Q \gamma_{\nu} q 
-  \partial_\nu \overline Q \gamma_{\mu} q\right)
\Bigr] ,
\eea 
where $a m_{q\{Q\}} = (1/\kappa_{q\{Q\}} - 1/\kappa_{cr})/2$, and $\overline m = (m_q + m_Q)/2$. 
The renormalisation constant, $Z_T^{(0)}(g_0^2,\mu)$ and the operator improvement coefficients, $b_T(g_0^2)$ and $c_T(g_0^2)$, are specified in table~\ref{table1}, where we also 
give the basic information about our lattices. 
\begin{table}[h!!]
\centering 
\begin{tabular}{cc}  \hline \hline
{\phantom{\huge{l}}}\raisebox{-.2cm}{\phantom{\Huge{j}}}
{\underline{\sc Set 1}} & $24^3\times 64$, $\beta=6.2$, $c_{SW}=1.614$ \\ \hline 
{\phantom{\huge{l}}}\raisebox{-.2cm}{\phantom{\Huge{j}}}
&  200 configs;\ $a^{-1}=2.7(1)$~GeV;\ $t_x=27$  \\
{\phantom{\huge{l}}}\raisebox{-.2cm}{\phantom{\Huge{j}}}
$\kappa_Q$&  0.125;\ 0.122;\ 0.119;\ 0.115  \\
{\phantom{\huge{l}}}\raisebox{-.2cm}{\phantom{\Huge{j}}}
$\kappa_q$&  0.1344;\ 0.1349;\ 0.1352  \\
{\phantom{\huge{l}}}\raisebox{-.2cm}{\phantom{\Huge{j}}}
& $Z^{(0)}_T=0.876(2)$;\ $b_T^{\rm bpt}=1.22$;\ $c_T=0.06$  \\
\hline
{\phantom{\huge{l}}}\raisebox{-.2cm}{\phantom{\Huge{j}}}
{\underline{\sc Set 2}} & $24^3\times 64$, $\beta=6.2$, $c_{SW}=1.614$ \\ \hline 
{\phantom{\huge{l}}}\raisebox{-.2cm}{\phantom{\Huge{j}}}
&  200 configs;\ $a^{-1}=2.8(1)$~GeV;\ $t_x=31$  \\
{\phantom{\huge{l}}}\raisebox{-.2cm}{\phantom{\Huge{j}}}
$\kappa_Q$&  0.128;\ 0.125;\ 0.122;\ 0.119;\ 0.116  \\
{\phantom{\huge{l}}}\raisebox{-.2cm}{\phantom{\Huge{j}}}
$\kappa_q$&  0.1344;\ 0.1346;\ 0.1348;\ 0.1350;\ 0.1352  \\
{\phantom{\huge{l}}}\raisebox{-.2cm}{\phantom{\Huge{j}}}
& $Z^{(0)}_T=0.876(2)$;\ $b_T^{\rm bpt}=1.22$;\ $c_T=0.06$  \\
\hline
{\phantom{\huge{l}}}\raisebox{-.2cm}{\phantom{\Huge{j}}}
{\underline{\sc Set 3}} & $32^3\times 70$, $\beta=6.45$, $c_{SW}=1.509$ \\ \hline 
{\phantom{\huge{l}}}\raisebox{-.2cm}{\phantom{\Huge{j}}}
&  100 configs;\ $a^{-1}=3.8(1)$~GeV;\ $t_x=34$  \\
{\phantom{\huge{l}}}\raisebox{-.2cm}{\phantom{\Huge{j}}}
$\kappa_Q$&  0.1285;\ 0.125;\ 0.122;\ 0.119;\ 0.116;\ 0.114  \\
{\phantom{\huge{l}}}\raisebox{-.2cm}{\phantom{\Huge{j}}}
$\kappa_q$&  0.1349;\ 0.1351;\ 0.1352;\ 0.1353  \\
{\phantom{\huge{l}}}\raisebox{-.2cm}{\phantom{\Huge{j}}}
& $Z^{(0)}_T=0.883(2)$;\ $b_T^{\rm bpt}=1.20$;\ $c_T^{\rm bpt}=0.02$ \\
\hline\hline
\end{tabular}
%%%%%%%%%%%%%%
{\caption{\small\sl \label{table1} Details on the lattices used in this work 
including the values of  
the Wilson heavy ($\kappa_Q$) and light ($\kappa_q$) mass parameters; 
${\cal O}(a)$ improvement coefficient of the action $c_{SW}$~\cite{cSW} and of the 
tensor current $c_T$~\cite{gupta}, $b_T$; 
renormalisation constant $Z_{T}^{(0)}(1/a)$~\cite{Znostre}. When 
a nonperturbative value is not available, we take its estimate in 
boosted perturbation theory (``bpt").}}
\end{table}
By using the standard definition of the vector current form factor 
\bea \label{f5}
\langle V(p^\prime,\lambda )\vert V^\mu(0) \vert B(p)\rangle = 
 \epsilon^{\mu \nu \alpha \beta} 
e^{\ast}_\nu(p^\prime,\lambda) p_\alpha p^{\prime}_\beta\ {2\ V(q^2) \over M_B +m_V} \;,
\eea
one can easily see that the improvement of the bare tensor current leaves 
the form factor $T_2(q^2)$ unchanged, whereas the form factor 
$T_1(q^2)$ is modified as
\bea
T_1^{\rm impr.}(q^2) = T_1(q^2) - a c_T  {q^2\over M_B+m_V} V(q^2)\;.
\eea

To study the form factors' $q^2$-dependence we considered the following 
$12$ combinations of $\vec p_H$, and $\vec q$:
\bea
\label{momenta}
{{\vec p_H =\left( 0,0,0\right)}}&\&&
\vec q \in \big\{ \left( 0,0,0\right)_{1}; \left( 1,0,0\right)_{4}; \left( 1,1,0\right)_{6}; 
\left( 1,1,1\right)_{4}; \left( 2,0,0\right)_{4} \big\},\\
&& \nonumber\\
\vec p_H =\left( 1,0,0\right)&\&&
\vec q \in \big\{ \left( 0,0,0\right)_{4}; \left( 0,1,0\right)_{12}; \left(0,1,1\right)_{6};  
\left( 1,0,0\right)_{4}; \left( 1,1,0\right)_{12};
\left( 1,1,1\right)_{12};\left( 2,0,0\right)_{4} \big\}, \nonumber
\eea
in units of $(2 \pi/La)$, the elementary momentum on the lattice with 
periodic boundary conditions. The index after each parenthesis 
in~(\ref{momenta}) denotes the number of independent correlation 
functions $C^{(3)}_{\zeta\mu\nu}(t_x,t_y;\vec q, \vec p_H)$, for a given combination of $\vec p_H$ and $\vec q$.  Those are deduced 
after  applying the symmetries: parity, charge conjugation, and the discrete 
cubic rotations. The plateaux of the ratios 
(\ref{ratio}) are typically found for $t_y\in [10,15]$.~\footnote{
Even after applying the available symmetries to the problem in hands, for each 
combination of $\kappa_Q$-$\kappa_q$ we still have 
$73$ correlation functions $C^{(3)}_{\zeta\alpha\beta}(t_x,t_y)$ when running over the 
ensemble of momenta~(\ref{momenta}). That means inspecting $4453$ ratios~(\ref{ratio}) 
and from the corresponding plateaux we extracted $671$ values for $T_{1,3}(q^2)$, and 
$732$ values of $T_2(q^2)$ form factor. We decided not to insert such formidable tables  
of numbers in this paper. A reader interested in those numbers can obtain them upon request 
from the authors.} 
The form factors $T_{1,2,3}(q^2)$ are then extracted by minimising the $\chi^2$ on the  corresponding set of plateaux of (\ref{ratio}). 
When both mesons are at rest only the form factor $T_2$ can be computed, whereas in 
other kinematical situations we obtain all $3$ form factors. In the following we will 
focus on $T_1$ and $T_2$.

\section{$T^{B\to K^\ast}(0)$ and $T^{B\to K^\ast}(0)/T^{B\to \rho}(0)$}

\subsection{Extrapolating to $B$ at $q^2=0$\label{ss1}}

As we already mentioned, in our lattice study we can extract the form factors 
at $q^2=0$, for each combination of $\kappa_Q$-$\kappa_q$, in all three of our datasets 
(see table~\ref{table1}). The form factors that we directly 
compute on the lattice cover a range of $q^2$'s straddling 
around zero, so that either one of the kinematical configurations~(\ref{momenta}) 
coincides with  $q^2=0$, or we have to interpolate the form factors calculated 
in the vicinity of $q^2=0$ to $q^2=0$. In the latter case the results are insensitive 
to the interpolation formula used.~\footnote{To check the insensitivity to 
the interpolation formula we used the forms discussed in 
eqs.~(\ref{bk1},\ref{bk2}) of the present paper, in addition to the pole/dipole form, i.e.,
$T_1(q^2) = T(0)/(1-q^2/m_1^2)^2$, $T_2(q^2) = T(0)/(1-q^2/m_2^2)$. }

A smooth linear mass interpolation (extrapolation) is needed to 
reach the $H\to {K^\ast}$ ($H\to\rho$) form factor, where $H$ is our fictitious 
heavy-light meson that is accessible from our lattice. This is done by fitting to 
\bea
T_1^{H\to V}(q^2=0) = T_2^{H\to V}(q^2=0)\equiv 
T^{H\to V}(0) = \alpha_H + \beta_H m_P^2\,,
\eea
where $m_P$ is the light pseudoscalar meson, while $\alpha_H$ and $\beta_H$ are the fit parameters. $T^{H\to K^\ast}(0)$ ($T^{H\to \rho}(0)$) is then obtained after choosing $m_P=m_K^{\rm phys.}$ ($m_\pi^{\rm phys.}$), where $m_K(\pi)^{\rm phys.}$ are identified on the lattice by using the method of physical lattice planes~\cite{giusti}. 
Such obtained values for $T(0)$, together with the masses in physical units, 
are given in table~\ref{table2}. We also list our results for 
$T^{H\to K^\ast}(0)/T^{H\to \rho}(0)$, which are simply obtained
as $(1 + m_K^{\rm phys.}\beta_H/\alpha_H)/(1 + m_\pi^{\rm phys.}\beta_H/\alpha_H)$.

\begin{figure}
\begin{center}
\begin{tabular}{@{\hspace{-0.7cm}}c}
\epsfxsize10.2cm\epsffile{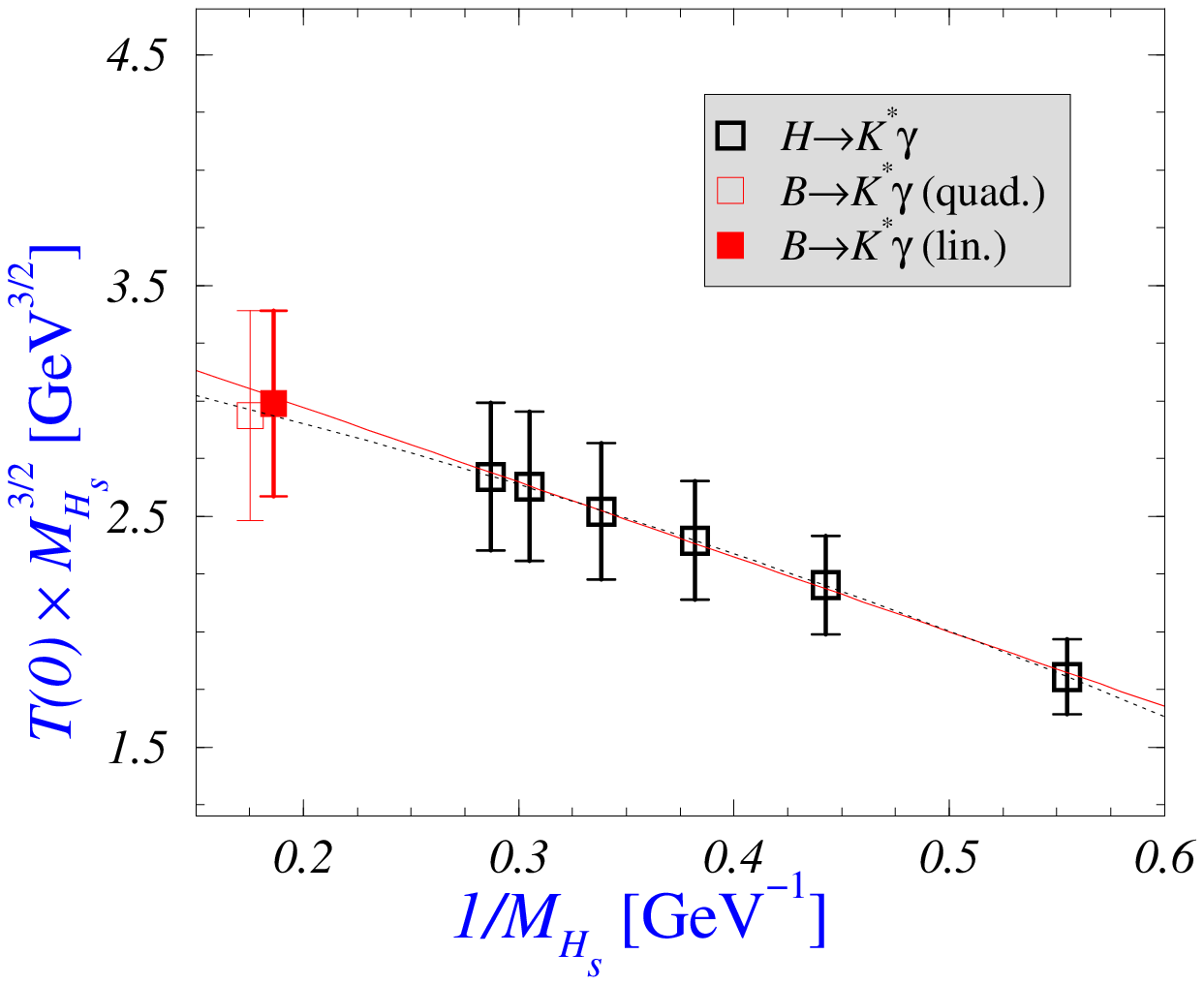}   \\
\epsfxsize10.2cm\epsffile{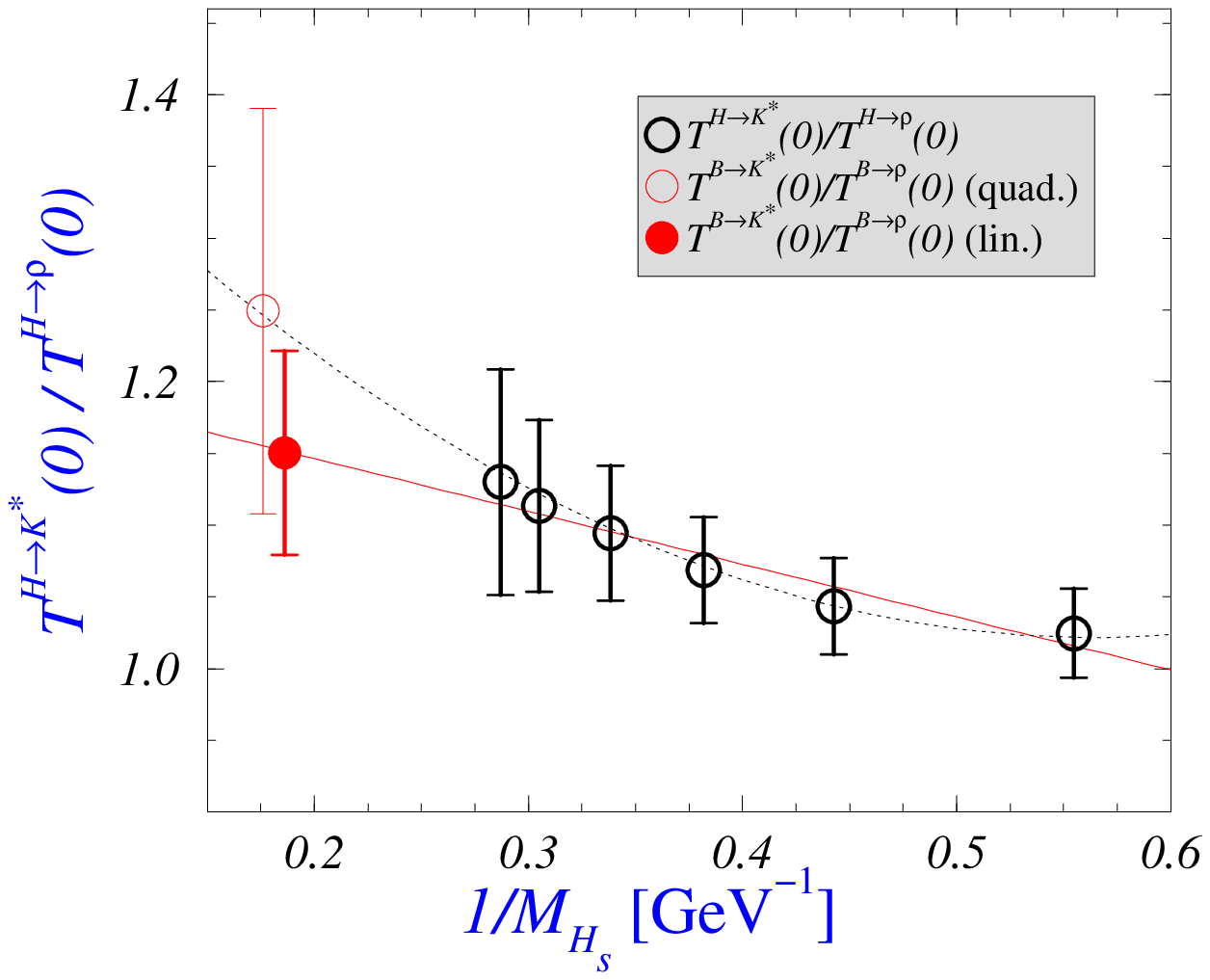}   \\
\end{tabular}
%%%%%%%%%%%%%%%%%%%%%%%%%%%%%%%%%%%%%%%%%%%%%%%%%%%%%%%%%%%%%%%%%%
\caption{\label{fig1}{\small \sl In the upper plot we show the extrapolation of the $H\to K^\ast\gamma$ form factor (multiplied by $M_{H_s}^{3/2}$) from $1/M_{H_s}$, directly accessible on the lattice at $\beta =6.45$, 
to $1/M_{B_s}$. Linear and quadratic fit to the data are denoted by the full and dotted lines respectively. The result of the quadratic extrapolation (empty square) is slightly shifted to left to 
make it discernible from the linear extrapolation result (filled square). The equivalent situation for the SU(3) breaking effect is shown in the lower plot.}}
%%%%%%%%%%%%%%%%%%%%%%%%%%%%%%%%%%%%%%%%%%%%%%%%%%%%%%%%%%%%%%%%%%
\end{center}
\end{figure}

\begin{table}[h!]
\centering
\begin{tabular}{|c|c|c|c|c|c|} \cline{2-6} 
 \multicolumn{1}{c|}{ \raisebox{.5cm}[8mm][5mm] }  
 & { $\kappa_Q$}  &  {$M_{H_s}$ [GeV]}  &  { $T^{H\to K^\ast}(0)$} & { $T^{H\to \rho}(0)$}& {$\displaystyle{
\frac{T^{H\to K^\ast}(0)}{T^{H\to \rho}(0)}}$} \\   \hline
{\phantom{\Large{l}}}\raisebox{.2cm}{\phantom{\Large{j}}}
{\sc Set 1} & 0.125 & 1.79(5) &  0.74(5) &  0.70(6) & 1.06(3)  \\
{\phantom{\Large{l}}}\raisebox{.2cm}{\phantom{\Large{j}}}
            & 0.122 & 2.05(5) &  0.70(5) &  0.66(7) &  1.07(3) \\
{\phantom{\Large{l}}}\raisebox{.2cm}{\phantom{\Large{j}}}
            & 0.119 & 2.29(6) &  0.65(6) &  0.60(8) &  1.09(4) \\
{\phantom{\Large{l}}}\raisebox{.2cm}{\phantom{\Large{j}}}
            & 0.115 & 2.59(7) &  0.60(7) &  0.55(9) &  1.10(6) \\ \hline
{\phantom{\Large{l}}}\raisebox{.2cm}{\phantom{\Large{j}}}
{\sc Set 2} & 0.128 & 1.57(4) &  0.80(11) &  0.75(16) & 1.05(5) \\
{\phantom{\Large{l}}}\raisebox{.2cm}{\phantom{\Large{j}}}
            & 0.125 & 1.87(6) &  0.77(8) &  0.73(12)& 1.06(4) \\
{\phantom{\Large{l}}}\raisebox{.2cm}{\phantom{\Large{j}}}
            & 0.122 & 2.13(7) &  0.72(7) &  0.68(10) & 1.05(5)\\
{\phantom{\Large{l}}}\raisebox{.2cm}{\phantom{\Large{j}}}
            & 0.119 & 2.39(7) &  0.67(7) &  0.63(10) & 1.06(7)\\
{\phantom{\Large{l}}}\raisebox{.2cm}{\phantom{\Large{j}}}
            & 0.116 & 2.62(8) &  0.62(8) &  0.57(11) & 1.09(9)\\ \hline
{\phantom{\Large{l}}}\raisebox{.2cm}{\phantom{\Large{j}}}
{\sc Set 3} & 0.1285 & 1.80(6) &  0.75(7) &  0.72(10)& 1.02(3) \\
{\phantom{\Large{l}}}\raisebox{.2cm}{\phantom{\Large{j}}}
            & 0.125 & 2.26(7) &  0.65(6) &  0.61(9) & 1.04(3)\\
{\phantom{\Large{l}}}\raisebox{.2cm}{\phantom{\Large{j}}}
            & 0.122 & 2.62(9) &  0.57(5) &  0.51(7)& 1.07(4) \\
{\phantom{\Large{l}}}\raisebox{.2cm}{\phantom{\Large{j}}}
            & 0.119 & 2.96(10) &  0.50(5) &  0.43(7) & 1.09(5)\\
{\phantom{\Large{l}}}\raisebox{.2cm}{\phantom{\Large{j}}}
            & 0.116 & 3.28(11) &  0.44(5) &  0.38(7)& 1.11(6) \\
{\phantom{\Large{l}}}\raisebox{.2cm}{\phantom{\Large{j}}}
            & 0.114 & 3.48(12) &  0.42(4) &  0.35(7)& 1.13(8) \\ \hline
\end{tabular}
\vspace*{0.4cm}
\caption{\label{table2}\small\sl The form factors at $q^2=0$ computed directly on the 
lattice at a fixed value of the heavy quark and for all  of our datasets. }
\end{table}
To extrapolate in the heavy quark mass we then use the heavy quark scaling 
law which tells us that $T^{H\to V}(0)\times m_Q^{3/2}$ should scale as 
a constant, up to corrections proportional to $1/m_Q^n$. Instead of 
the heavy quark mass, we may take the mass of the corresponding heavy-light meson consisting 
of a heavy $Q$-quark and the light $s$-quark.  The reason for using 
the strange light quark is that it is directly accessible on our lattices 
whereas for the light $u/d$-quark one needs to make an extrapolation 
which increases the error on the heavy-light meson mass.  
In other words we fit our data to
\bea\label{eq9}
T^{H\to V}(0)\times M_{H_s}^{3/2} = c_0 + { c_1\over M_{H_s} } + { c_2\over {M_{H_s}}^2 }\,, 
\eea
where $c_{0,1,2}$ are the fit parameters. 
From the plot in fig.~\ref{fig1} we see a pronounced linear behavior in 
$1/{M_{H_s}}$, 
%%%%%%%%%%%%%%%%%%%%%%%%%%%%%%%%%%%%%%%%%%%%%%%%%%%%%%%%%%%
%%%%%%%%%%%%%%%%%%%%%%%%%%%%%%%%%%%%%%%%%%%%%%%%%%%%%%%%%%%
which is why we will take the result of the linear extrapolation ($c_2=0$)  
as our main result.  
As it could be guessed from fig.~\ref{fig1}, at $\beta=6.45$, 
the extrapolated value does not change if we leave out from the fit 
the point corresponding to the lightest of our heavy quarks. 
Since we have more (and heavier) masses at $\beta=6.45$, we prefer to quote 
the results obtained from that dataset (Set 3), namely,
\bea\label{F12}
T^{B\to K^\ast}_{\rm lin.}(0)= 0.24(3)\,,\qquad T^{B\to K^\ast}_{\rm quad.}(0) = 0.23(4)\,,
\eea
where ``lin." and ``quad." stand for the linearly and quadratically extrapolated 
form factors to $1/M_{B_s}$. 
The results of the strategy discussed in this subsection for all our 
lattices are listed in table~\ref{table3}. 

\begin{table}[h]
\centering
\begin{tabular}{|c|c|c|c|c|} \cline{2-5}
 \multicolumn{1}{c|}{ \raisebox{.5cm}[8mm][5mm]}  & { $T^{B\to K^\ast}_{\rm lin.}(0)$} & { $T^{B\to K^\ast}_{\rm quad.}(0)$} & 
 { $T^{B\to K^\ast}_{\rm lin.}(0)/T^{B\to \rho}_{\rm lin.}(0)$} & { $T^{B\to K^\ast}_{\rm quad.}(0)/T^{B\to \rho}_{\rm quad.}(0)$}  \\ \hline 
{\phantom{\Large{l}}}\raisebox{.2cm}{\phantom{\Large{j}}}
{\sc Set 1} & 0.25(3) & 0.28(7) &  1.14(11) &  1.24(22) \\ \hline
{\phantom{\Large{l}}}\raisebox{.2cm}{\phantom{\Large{j}}}
{\sc Set 2} & 0.28(6) & 0.29(9) &  1.08(12) &  1.17(31) \\ \hline
{\phantom{\Large{l}}}\raisebox{.2cm}{\phantom{\Large{j}}}
{\sc Set 3} & 0.24(3) & 0.23(4) &  1.15(7) &  1.25(14) \\
\hline \end{tabular}
\vspace*{0.4cm}
\caption{\label{table3}\small\sl The form factors resulting from the linear and 
quadratic extrapolation~(\ref{eq9}) as obtained from all of our 3 datasets specified 
in table~\ref{table1}.}
\end{table}
As an illustration, the linear fit with our data at $\beta=6.45$ gives
\bea
T^{B\to K^\ast}(0) = {3.6(7)\ \gev^{3/2}\over M_{B_s}^{3/2} } \times \left[ 1 - {0.9(1)\ \gev\over M_{B_s}} \right]\,. 
\eea
The slope in $1/M_{B_s}$ is very close to what has been observed  
in the lattice studies of the heavy-light decay constants~\cite{heavy}, 
and of the $B\to \pi$ semileptonic decay 
form factor (see eq.~(19) in ref.~\cite{our_BPI}).

From table~\ref{table3} we see that the ratio of 
$B\to K^\ast$ and $B\to \rho$ form factors has a large error. 
This error comes from $T^{B\to \rho}(0)$, and in particular from the  
light mass extrapolation of the form factors to reach $T^{H\to \rho}(0)$. That error 
is larger for larger $m_H$, which gets further inflated after extrapolating to $B$. 
In contrast, the extrapolation to reach $T^{H\to K^\ast}(0)$ is not needed as the
$K^\ast$ mass falls in the range of the vector meson masses that are  directly 
simulated on our lattices. When extrapolated to $B$-meson, the SU(3) breaking ratio 
of the form factor has a large error and is very sensitive to the inclusion 
of the quadratic term in the extrapolation.

Before closing this subsection, let us also mention that, contrary to HQET, 
one cannot match the short distance behaviour of our QCD results to the soft 
collinear effective theory (SCET) in which the $T^{H\to V}(0)\times m_Q^{3/2}$ 
scaling law is manifest. This is still an unsolved theoretical problem 
and hopefully a recent development based on 
ref.~\cite{stewart} will help solving it. 
We note however that the inclusion of the matching of the tensor current anomalous 
dimension of QCD with HQET  produces a numerically marginal effect ($1\div 2\%$ on 
the central values). 
We hope that the similar will hold once such a matching of QCD with SCET 
becomes possible.

\subsection{Extrapolating to $B$ at $q^2\neq 0$ and then to $q^2=0$\label{ss2}}

\begin{figure}
\vspace*{-.4cm}
\begin{center}
\begin{tabular}{@{\hspace{.1cm}}c}
\epsfxsize11.4cm\epsffile{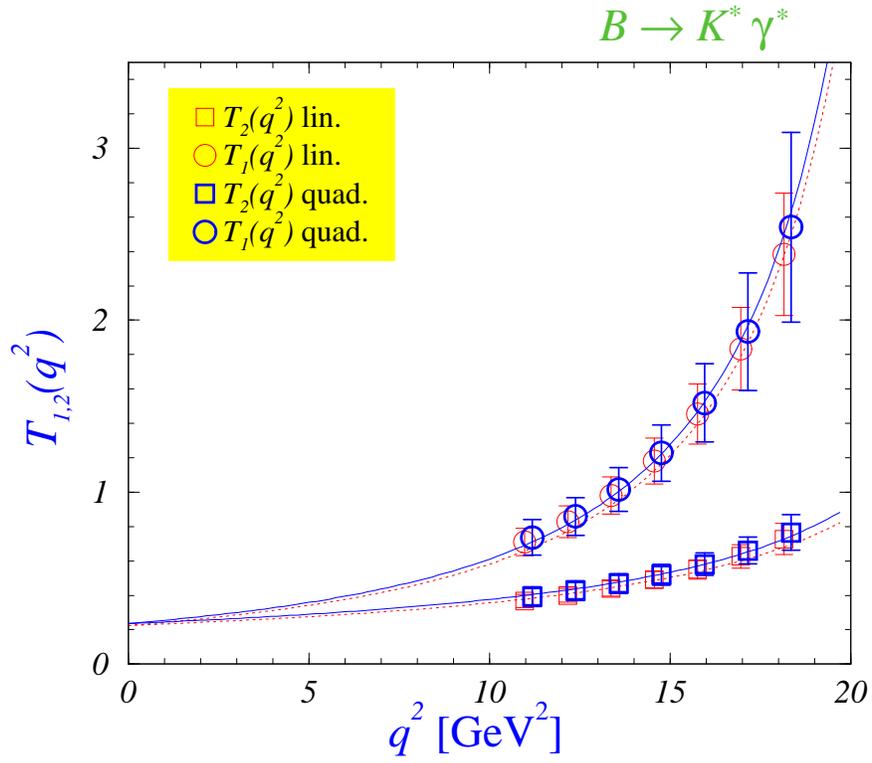}   \\
\end{tabular}
%%%%%%%%%%%%%%%%%%%%%%%%%%%%%%%%%%%%%%%%%%%%%%%%%%%%%%%%%%%%%%%%%%
\caption{\label{fig2}{\small\sl The form factors $T_{1,2}(q^2)$  
relevant for $B\to K^\ast \gamma^\ast$ decay, obtained after extrapolating (linearly and quadratically) our 
data at $\beta=6.45$ in inverse heavy meson mass. Also shown are the curves 
fitting the $q^2$ dependence to the expressions given in eqs.~(\ref{bk1},\ref{bk2}). }}
%%%%%%%%%%%%%%%%%%%%%%%%%%%%%%%%%%%%%%%%%%%%%%%%%%%%%%%%%%%%%%%%%%
\end{center}
\end{figure}

To check on the results obtained in the previous subsection we now also 
employ the standard method and extrapolate our results at fixed $v\ndot p^\prime$ 
to $M_B$. The main assumption here is that the HQET scaling laws are valid for   
all our $v\ndot p^\prime$, i.e., not only for those that are very small  
compared to the heavy meson mass. 
\begin{itemize}
\item From our directly accessed masses and $q^2$'s, one identifies  
$v\ndot p^\prime =(M_H^2 + m_V^2-q^2)/2M_H$, where $H$ is the heavy-light meson and 
$V$ stands for either $K^\ast$ or $\rho$. 
In physical units, the range of available $v\ndot p^\prime$ 
is nearly equal for all of our lattices, namely 
$0.9~\gev \lesssim v\ndot p^\prime \lesssim  1.8~\gev$. We emphasize that the kinematical configurations which we were able to explore are those listed in eq.~(\ref{momenta}). 
Proceeding like in ref.~\cite{our_BPI}, we chose 5, 6, 7 equidistant $v \ndot p^\prime$ points 
for our dataset 1, 2, 3,  respectively.  
The form factors, $T_{1,2}(v\ndot p^\prime)$, are then linearly interpolated (extrapolated) 
to $m_{K^\ast}$ ($m_\rho$) for each of our heavy quarks. 
\item We construct
\bea \label{phis}
\Phi_1(M_H,v\ndot p^\prime) = w(M_H){ T_1(v\ndot p^\prime)\over \sqrt{\ M_H\ }}\,,\quad
\Phi_2(M_H,v\ndot p^\prime) = w(M_H) T_2(v\ndot p^\prime)\times \sqrt{\ M_H\ },
\eea
which, in HQET, are expected to scale as constants, up to corrections $\propto 1/M_H^n$. 
Fitting our data to 
\bea\label{extrap}
\Phi_{1,2}(M_H,v\ndot p^\prime) = d_0(v\ndot p^\prime) + {d_1(v\ndot p^\prime)\over 
M_H} + {d_2(v\ndot p^\prime)\over M_H^2}\,, 
\eea 
either linearly ($d_2=0$) or quadratically ($d_2\neq 0$), we can extrapolate  
to the $B$-meson mass. Since we use the scaling law which is manifest 
in HQET the factor $w(M_H)$ in eq.~(\ref{phis}) 
accounts for the mismatch of the leading order anomalous dimensions 
in QCD ($\gamma_T= 8/3$) and in HQET ($\widetilde \gamma_T= -4$) for the 
tensor density, namely, 
\bea
w(M_H) = \left( {\alpha_s(M_H)\over \alpha_s(M_B)}
\right)^{- {\gamma_T - \widetilde \gamma_T\over 2 \beta_0} } \times 
 \left( {\alpha_s(1/a)\over \alpha_s(M_H)}
\right)^{- {\gamma_T\over 2 \beta_0} } \left[ 1 - J_T { \alpha_s(1/a) -  \alpha_s(M_H) \over 4 \pi}\right]\,.
\eea
The numerator in the first factor match our QCD form factors with their HQET counterparts, 
while the dominator does the opposite to the result of the extrapolation to $M_B$. 
The second factor, instead, provides the NLO evolution 
from $\mu=1/a$ to $\mu=M_H$. For $N_f=0$, $\beta_0=11$, and $J_T=2.53$.  
\item The extrapolation~(\ref{extrap}) is made both linearly and quadratically. 
The differences between the corresponding results are essentially indistinguishable, 
as it can be seen in fig.~\ref{fig2} where we plot the results for 
$B\to K^\ast\gamma^\ast$ form factors ($\gamma^\ast$ stands for the off-shell photon) 
obtained at $\beta = 6.45$ (Set 3). These and the similar results we obtained 
at $\beta=6.2$ are  
collected in table~\ref{table4}.
 
\end{itemize}
%%%%%%%%%%%%%%%%%%%%%%%%%%%%%%%%%%%%%%%%%%%%%%%%%%%%%%%%%%%
%%%%%%%%%%%%%%%%%%%%%%%%%%%%%%%%%%%%%%%%%%%%%%%%%%%%%%%%%%%
%%%%%%%%%%%%%%%%%%%%%%%%%%%%%%%%%%%%%%%%%%%%%%%%%%%%%%%%%%%
\begin{table}
\hspace*{4cm}
\begin{center}
\begin{tabular}{|c|c|c|c|c|c|c|} 
\multicolumn{3}{c}{\underline{\sc Set 1}}&\multicolumn{1}{c}{\raisebox{-.8cm} }&\multicolumn{3}{c}{\underline{
\sc Set 2}}\\
\cline{1-3}  \cline{5-7} 
\hspace{-4.mm}{\phantom{\huge{l}}}\raisebox{-.2cm}{\phantom{\Huge{j}}}
{  $q^2\ [{\rm GeV}^2]$\hspace{3mm}}& { \hspace{-1mm}$T_1^{B\to K^\ast}(q^2)$\hspace{1mm}} &{\, \ $T_2^{B\to K^\ast}(q^2)$}&$\qquad$ &{  $q^2\ [{\rm GeV}^2]$\hspace{3mm}}& { \hspace{-1mm}$T_1^{B\to K^\ast}(q^2)$\hspace{1mm}} &{\, \ $T_2^{B\to K^\ast}(q^2)$}\\ 
\cline{1-3}  \cline{5-7}  
{\phantom{\Large{l}}}\raisebox{.2cm}{\phantom{\Large{j}}}
{ 12.2} & {$\mathsf{0.76(24)^{+.00}_{-.02}}$} & {$\mathsf{0.50(13)^{+.00}_{-.07}}$} & &
{ 13.5} & {$\mathsf{0.70(17)^{+.13}_{-.00}}$} & {$\mathsf{0.52(9)^{+.00}_{-.14}}$} \\
{\phantom{\Large{l}}}\raisebox{.2cm}{\phantom{\Large{j}}}
{ 13.6} & {$\mathsf{0.94(22)^{+.00}_{-.06}}$} & {$\mathsf{0.54(12)^{+.00}_{-.07}}$} & &
{ 14.4} & {$\mathsf{0.81(18)^{+.11}_{-.00}}$} & {$\mathsf{0.56(9)^{+.00}_{-.15}}$} \\
{\phantom{\Large{l}}}\raisebox{.2cm}{\phantom{\Large{j}}}
{ 15.1} & {$\mathsf{1.19(19)^{+.00}_{-.12}}$} & {$\mathsf{0.60(10)^{+.00}_{-.07}}$} & &
{ 15.3} & {$\mathsf{0.95(19)^{+.07}_{-.00}}$} & {$\mathsf{0.61(9)^{+.00}_{-.16}}$} \\
{\phantom{\Large{l}}}\raisebox{.2cm}{\phantom{\Large{j}}}
{ 16.5} & {$\mathsf{1.53(17)^{+.00}_{-.21}}$} & {$\mathsf{0.64(6)^{+.00}_{-.07}}$} & &
{ 16.1} & {$\mathsf{1.11(21)^{+.03}_{-.0}}$} & {$\mathsf{0.66(9)^{+.00}_{-.15}}$} \\
{\phantom{\Large{l}}}\raisebox{.2cm}{\phantom{\Large{j}}}
{ 17.9} & {$\mathsf{2.03(21)^{+.00}_{-.37}}$} & {$\mathsf{0.72(2)^{+.00}_{-.07}}$}  & &
{ 17.0} & {$\mathsf{1.32(26)^{+.00}_{-.04}}$} & {$\mathsf{0.72(9)^{+.00}_{-.15}}$}\\
{\phantom{\Large{l}}}\raisebox{.2cm}{\phantom{\Large{j}}}
{ --} & {$\mathsf{-}$} & {$\mathsf{-}$}  & &
{ 17.8} & {$\mathsf{1.56(36)^{+.00}_{-.12}}$} & {$\mathsf{0.80(9)^{+.00}_{-.15}}$}\\
\cline{1-3}  \cline{5-7} 
\end{tabular}\\
\vspace*{12mm}
\begin{tabular}{|c|c|c|} 
\multicolumn{3}{c}{\underline{\sc Set 3}\raisebox{-.8cm}} \\ 
 \hline
\hspace{-4.mm}{\phantom{\huge{l}}}\raisebox{-.2cm}{\phantom{\Huge{j}}}
{  $q^2\ [{\rm GeV}^2]$\hspace{3mm}}& { \hspace{-1mm}$T_1^{B\to K^\ast}(q^2)$\hspace{1mm}} &{\, \ $T_2^{B\to K^\ast}(q^2)$}\\ 
\hline   
{\phantom{\Large{l}}}\raisebox{.2cm}{\phantom{\Large{j}}}
{ 11.2} & {$\mathsf{0.73(10)^{+.00}_{-.02}}$} & {$\mathsf{0.39(6)^{+.00}_{-.02}}$} \\
{\phantom{\Large{l}}}\raisebox{.2cm}{\phantom{\Large{j}}}
{ 12.4} & {$\mathsf{0.86(11)^{+.00}_{-.03}}$} & {$\mathsf{0.43(6)^{+.00}_{-.03}}$} \\
{\phantom{\Large{l}}}\raisebox{.2cm}{\phantom{\Large{j}}}
{ 13.6} & {$\mathsf{1.02(13)^{+.00}_{-.04}}$} & {$\mathsf{0.47(6)^{+.00}_{-.03}}$} \\
{\phantom{\Large{l}}}\raisebox{.2cm}{\phantom{\Large{j}}}
{ 14.7} & {$\mathsf{1.23(16)^{+.00}_{-.05}}$} & {$\mathsf{0.52(6)^{+.00}_{-.03}}$} \\
{\phantom{\Large{l}}}\raisebox{.2cm}{\phantom{\Large{j}}}
{ 16.0} & {$\mathsf{1.52(23)^{+.00}_{-.07}}$} & {$\mathsf{0.58(7)^{+.00}_{-.03}}$} \\
{\phantom{\Large{l}}}\raisebox{.2cm}{\phantom{\Large{j}}}
{ 17.2} & {$\mathsf{1.93(34)^{+.00}_{-.10}}$} & {$\mathsf{0.66(8)^{+.00}_{-.03}}$} \\
{\phantom{\Large{l}}}\raisebox{.2cm}{\phantom{\Large{j}}}
{ 18.3} & {$\mathsf{2.54(55)^{+.00}_{-.16}}$} & {$\mathsf{0.76(10)^{+.00}_{-.03}}$} \\
  \hline 
\end{tabular}
%%%%%%%%%%%%%%%%%%%%%%%%%%%%%%%%%%%%%%%%%%%%%%%%%%%%%%%%%%%%%%%%%%
%\vspace*{-.75cm}
\caption{\label{table4}{\small\sl The values of the $B\to K^\ast\gamma^\ast$ form factors 
at several values of $q^2$. 
The first error in each result is statistical and the second is the difference between 
the results of linear and quadratic extrapolation in $1/M_H$ to $1/M_B$.}}
\end{center}
\end{table}
To reach the physically interesting case of the photon on-shell one needs to assume some functional dependence of the form factors and extrapolate the results of table~\ref{table4} 
down to $q^2=0$. It is very easy to convince oneself that the form factors $T_1$ and $T_2$ satisfy the  constraints very similar to those that govern the shapes of $F_+$ and $F_0$ semileptonic heavy to light pseudoscalar form factors. More specifically:
\begin{itemize}
\item[$\circ$] The nearest pole in the crossed channel, $J^P=1^-$, which contributes 
to the form factor $T_1$, is $M_{B^\ast_s}=5.42$~GeV in $B\to K^\ast$ transition. 
The form factor $T_2$, instead, receives the contribution from heavy $J^P=1^+$ resonances and multiparticle states both below and above the cut $(M_B+m_V)^2$. 
\item[$\circ$] HQET, which is relevant to the region of large $q^2$'s, 
suggests that the form factors scale with heavy quark/meson mass as 
$T_1 \sim \sqrt{M}$ and $T_2 \sim 1/\sqrt{M}$~\cite{isgur}, and therefore both form factors cannot be fit to the pole-like shapes.
\item[$\circ$] For the high energy region of the light meson 
in the rest frame of the heavy ($q^2\to 0$),  
we also have the scaling laws $T_1(E)\sim \sqrt{M}/E^2\sim M^{-3/2}$. Similar holds for 
the $T_2(E)$ form factor, i.e., both form factors scale as $M^{-3/2}$~\cite{jerome,SCET}. 
Moreover, the two are related via
\bea
T_1(E) = {M\over 2 E} T_2(E)\,.
\eea   
\end{itemize}
Thus the situation is analogous to the one in $B\to \pi\ell \nu$ decay, 
and we can use the simple parameterisation of ref.~\cite{BK}, 
\bea\label{bk1}
T_1(q^2) = {T(0)\over (1- \widetilde q^2) (1- \alpha \widetilde q^2)}\,,\qquad 
T_2(q^2) = {T(0)\over 1- \widetilde q^2/\beta }\,,
\eea
where $\widetilde q^2=q^2/M_{B^\ast_s}^2$. If we relax the $T_1/T_2 = {M/2 E}$ 
constraint, then a simple form~(\ref{bk1}) becomes 
\bea\label{bk2}
T_1(q^2) = {C_1\over 1- \widetilde q^2} + {C_2\over  1- C_3 \widetilde q^2 }\,,\qquad 
T_2(q^2) = {C_1+C_2 \over 1- C_4 \widetilde q^2  }\,.
\eea
Our data from table~\ref{table4} cannot distinguish between the two of the above 
parameterisations and in both cases we end up with the same value for $T(0)$. 
As in the previous subsection, as our final results we will quote 
those obtained at $\beta=6.45$, for which more and heavier mesons were accessible,
\bea\label{F2}
T^{B\to K^\ast}_{\rm lin.}(0)= 0.22(3)\,,\qquad T^{B\to K^\ast}_{\rm quad.}(0) = 0.24(4)\,.
\eea
The results for all three datasets are collected in table~\ref{table5}.
\begin{table}[h]
\centering
\begin{tabular}{|c|c|c|c|c|} \cline{2-5}
 \multicolumn{1}{c|}{ \raisebox{.5cm}[8mm][5mm]} & { $T^{B\to K^\ast}_{\rm lin.}(0)$} & { $T^{B\to K^\ast}_{\rm quad.}(0)$} & 
 { $T^{B\to K^\ast}_{\rm lin.}(0)/T^{B\to \rho}_{\rm lin.}(0)$} & { $T^{B\to K^\ast}_{\rm quad.}(0)/T^{B\to \rho}_{\rm quad.}(0)$}  \\   \hline
{\phantom{\Large{l}}}\raisebox{.2cm}{\phantom{\Large{j}}}
{\sc Set 1} & 0.25(6) & 0.28(10) &  1.2(3) &  1.3(7) \\ \hline
{\phantom{\Large{l}}}\raisebox{.2cm}{\phantom{\Large{j}}}
{\sc Set 2} & 0.23(6) & 0.23(6) &  1.1(3) &  1.1(2) \\ \hline
{\phantom{\Large{l}}}\raisebox{.2cm}{\phantom{\Large{j}}}
{\sc Set 3} & 0.22(3) & 0.24(4) &  1.3(2) &  1.3(3) \\
\hline \end{tabular}
\vspace*{0.4cm}
\caption{\label{table5}\small\sl The results of the extrapolation of the form factors from  
table~\ref{table4} assuming the $q^2$-dependence given in eqs.~(\ref{bk1},\ref{bk2}).}
\end{table}

\section{Final results and conclusion}

As our final results we will quote those obtained at $\beta =6.45$, 
since they have smaller discretisation errors. 
As central value we take the results obtained from the first method (subsec.~\ref{ss1}) 
which are given in table~\ref{table3} at $\mu=1/a$, in the (Landau) RI/MOM scheme. 
The matching to the $\msbar$ scheme is $1$ at 2-loop accuracy, but we still need to 
run our result from $\mu=1/a$ to $m_b$ which is made via
\bea
T^{B\to K^\ast}(0; m_b) = T(0;1/a) \left({\alpha_s(1/a)\over \alpha_s(m_b)}\right)^{-\gamma_T/2\beta_0} 
\left[ 1 - J_T {\alpha_s(1/a)- \alpha_s(m_b)\over 4\pi}\right]\,. 
\eea
For $N_f=0$ and $m_b=4.6(1)$~GeV, the running factor is very close to $1$, i.e.,
$0.99$ ($0.97$) for the form factors computed at $\beta=6.45$ ($\beta=6.2$). 
Our final result is
\bea\label{F3}
T^{B\to K^\ast}(q^2=0;\mu=m_b) = 0.24\pm 0.03^{+0.04}_{-0.01}\,. 
\eea
The spread of central values presented in table~\ref{table3} at $\beta =6.2$ (when multiplied 
by $0.97$) are used to attribute a systematic error to our final result~(\ref{F3}). 
The values in table~\ref{table5} are already at the scale $\mu = M_B \approx m_b$, and those given in  eq.~(\ref{F2}) are fully consistent with~(\ref{F3}). 
Our number is smaller when compared to the QCD sum rule results, 
\bea
T^{B\to K^\ast}(0) = 0.33(3)~\cite{LCSR},\ 0.38(6)~\cite{QCDSR}, 
\eea
although the recent progress show the tendency of lowering the ccentral value 
obtained by using LCSR, i.e., $T^{B\to K^\ast}(0) = 0.31(4)$~\cite{updateLCSR}. 
Concerning the previous lattice studies~\cite{lattice_old}, 
most of them were made at the time before the $T(0)M_H^{3/2}$-scaling 
law was known or before a significant statistical quality of the lattice 
data was feasible.

Our values for the ratio $T^{B\to K^\ast}(0)/T^{B\to \rho}(0)$, on the other hand, are
unstable, and we will only quote the average of 
the values obtained by using two methods discussed in this paper at $\beta =6.45$,  
\bea
T^{B\to K^\ast}(0)/T^{B\to \rho}(0) =1.2(1)\,, 
\eea
as our best estimate. This result agrees with the most recent LCSR estimate, 
$T^{B\to K^\ast}(0)/T^{B\to \rho}(0) =1.17(9)$~\cite{zwickyNEW}.

The method employed in this work relies on the use of a 
propagating heavy quark. In order to reach smaller values of 
$q^2$'s in an effective theory of heavy quark the
so called moving NRQCD has been developed, but the numerical 
quality of the signal does not appear very encouraging so far~\cite{mnrqcd}.  
Clearly, our method cannot be used to obtain a very accurate 
value of $T(0)$, mainly because the heavy quark extrapolations are involved. 
If we are to work with the physical $b$-quark mass on the lattice, 
we would need a very small lattice spacing, i.e., at least $a^{-1}\simeq 10$~GeV.  
For the volume corresponding to the lattice size $La=2$~fm 
that would require the simulations on the lattice with $100^3$ spatial points.   
Moreover, on the lattice with the periodic boundary conditions, the $q^2=0$ 
point is reached  when the energy of the vector meson (in the rest frame of the heavy) 
is
\bea
&&E^2 = m_{K^\ast}^2 + \left({2\pi\over La}\right)^2 
\vert \vec n\vert ^2 \stackrel{\downarrow}{=} \left({M_B^2+ m_{K^\ast}^2\over 2 M_B}\right)^2\cr 
&&\Rightarrow \vert \vec n\vert  = {M_B^2- m_{K^\ast}^2\over 4\pi M_B} \times La = (2.07\ {\rm fm}^{-1})\times La \,.
\eea
Therefore on a lattice of the size $La=2$~fm, one needs to give the  
kaon a momentum $ \vert \vec n\vert\approx 4$, for which it is 
highly unlikely to observe any signal of the correlation functions~(\ref{asym3}).  
Even if one organises the kinematics so that 
the momenta are shared between $B$- and $K^\ast$-mesons, the required spatial momenta 
are still too large for the reasonably accurate computation of  the correlation functions~(\ref{asym3}). Therefore, a progress in improving the quality of signals 
of correlation functions when the spatial momenta 
$\vert \vec n\vert > 1$ would be highly welcome. 
Summarising, as of now, a significant 
improvement on the precision of the $B\to K^\ast$ form factors does not look 
promising even in quenched approximation. 
The methodology of the extraction of the form factors 
can, however, be improved by combining the correlation functions~(\ref{asym3}) 
in double ratios, in a way similar to what  has recently been implemented 
in the lattice computation of heavy$\to$heavy~\cite{BD} and light$\to$light pseudoscalar meson decay form factors~\cite{Kpi}.

%\twocolumn

\end{document}